%% file: main.tex
\documentclass[letterpaper,5p,times,preprint]{elsarticle}
 
\usepackage[utf8]{inputenc}
\usepackage{siunitx}
\usepackage{graphicx}
\usepackage{amsmath,mathtools, cuted}
\usepackage{subcaption}
\usepackage[table,dvipsnames]{xcolor}
\usepackage{booktabs}
\usepackage{multirow}
\usepackage{bm}
\usepackage{textcomp}
\usepackage{float}
\usepackage{gensymb}
\usepackage{chemformula}
\usepackage{amsopn}
\usepackage{tabularx}
\usepackage{enumitem}
\usepackage{times}
\usepackage{lscape}
\usepackage{threeparttable}
\usepackage{tabularx}
\usepackage{caption}
\usepackage{pifont}
\usepackage{longtable}
\usepackage{threeparttablex}
\usepackage{listings}
\usepackage{hyperref}
\usepackage{cleveref}

\makeatletter
\def\ps@pprintTitle{%
 \let\@oddhead\@empty
 \let\@evenhead\@empty
 \def\@oddfoot{}%
 \let\@evenfoot\@oddfoot}
\makeatother

\crefname{appendix}{}{}

\hypersetup{
    hidelinks,
    colorlinks=false,
    linkcolor=black,
    citecolor=black,
    urlcolor=black
}

\begin{document}
\sloppy
\begin{frontmatter}

\title{Construction and Tuning of CALPHAD Models Using Machine-Learned Interatomic Potentials and Experimental Data: A Case Study of the Pt–W System}

\author[1]{Courtney Kunselman}
\ead{cjkunselman18@tamu.edu}

\author[1]{Siya Zhu}
\ead{siyazhu@tamu.edu}

\author[1]{Do\u{g}uhan Sar\i{}t\"urk}
\ead{sariturk@tamu.edu}

\author[1,2,3]{Raymundo Arr\'oyave}
\ead{raymundo.arroyave@tamu.edu}

\affiliation[1]{organization={Department of Materials Science and Engineering, Texas A\&M University},
                city={College Station},
                postcode={77843}, 
                state={TX},
                country={USA}}

\affiliation[2]{organization={J. Mike Walker '66 Department of Mechanical Engineering, Texas A\&M University},
                city={College Station},
                postcode={77843}, 
                state={TX},
                country={USA}}

\affiliation[3]{organization={Wm Michael Barnes '64 Department of Industrial and Systems Engineering, Texas A\&M University},
                city={College Station},
                postcode={77843}, 
                state={TX},
                country={USA}}

\begin{abstract}
This work introduces \texttt{PhaseForgePlus} -- a computationally efficient, fully open-source workflow for physically-informed CALPHAD model generation and parameter fitting. Using the Pt-W system as an example, we show that the integration of Machine Learning Potentials into the Alloy Theoretic Automated Toolkit can produce physically grounded Gibbs energy descriptions requiring only slight adjustments to produce accurate phase diagrams. Employing the Jansson derivative method in the context of experimental observations, such adjustments can be efficiently and robustly determined through gradient-informed optimization procedures.
\end{abstract}


\end{frontmatter}

\input{01_introduction}
\input{02_methods}
\input{03_results}

\input{04_conclusion}
\input{05_codeAvailability}
\input{06_acknowledgements}

\medskip
\bibliographystyle{unsrt}
\bibliography{ref}

\end{document}

%% file: 01_introduction.tex
\section{Introduction}
Alloy phase diagrams are foundational to understanding thermodynamic stability and guiding alloy design. The CALPHAD (\textit{CALculation of PHAse Diagrams}) method remains the standard approach for constructing these diagrams by parameterizing the Gibbs free energy of each phase---typically through polynomial expansions in composition and temperature. Model parameters are calibrated against experimental measurements and/or high-fidelity computational data to achieve the desired level of accuracy.

The study of phase diagrams through experiments is a traditional and reliable approach. However, experimental investigations are often costly and time-consuming, as only a limited subset of compositions and temperatures can be explored, and reaching thermodynamic equilibrium may require prolonged annealing. Existing thermodynamic databases, such as those developed by the Scientific Group Thermodata Europe (SGTE) or the Thermo-Calc High Entropy Alloy database (TCHEA)---are typically constructed based on experimental data and can be directly used for phase diagram calculations. However, these databases have important limitations: they may lack coverage of all relevant phases or element combinations, and many are commercial products developed through proprietary closed- source methodologies. This restricts transparency and makes the modeling process somewhat of a black box. Moreover, access to commercial CALPHAD tools and databases may be limited for some academic or industrial groups due to licensing and cost constraints.

On the other hand, CALPHAD models can be constructed purely from theoretical computations, using \textit{ ab initio} methods or machine learning-based interatomic potentials (MLIPs), with the aid of software packages such as the Alloy Theoretic Automated Toolkit (ATAT)~\cite{van2002alloy,van2017software}. In our previous work~\cite{zhu2025accelerating,zhu2025machinelearningpotentialsalloys}, we introduced \texttt{PhaseForge}~\cite{sariturk_2025_15730911}, a newly developed package that integrates MLIPs into the ATAT framework and enables the automated construction of CALPHAD-type thermodynamic models. Although such approaches rely on physically grounded estimates of energy and entropy, their accuracy may still be constrained by trade-offs in computational efficiency as well as modeling assumptions and approximations.

To address these accuracy concerns, the parameters of the thermodynamic models generated from the \textit{ ab initio} approaches---such as those employed by \texttt{PhaseForge}---can be refined using experimental data collected from the literature. In recent work, we demonstrated that the recently formalized Jansson derivative method~\cite{kunselman2024analytically} enables gradient-based optimization of CALPHAD model parameters~\cite{kunselman2025analytical} within the openly available \texttt{PyCalphad}~\cite{otis2017pycalphad} and \texttt{ESPEI}~\cite{bocklund2019espei} toolchain. Across four alloy systems studied, conjugate gradient optimization yielded computational efficiency improvements ranging from one to three orders of magnitude over \texttt{ESPEI}’s default black-box optimizer, which uses ensemble Bayesian inference through Markov Chain Monte Carlo (MCMC).

In this work, we present a workflow for constructing CALPHAD models using theoretically derived thermodynamic data, followed by gradient-informed refinement using experimental data from the literature. As a case study, we apply this workflow to the Pt--W binary system. The initial model is generated using MLIP-based thermodynamic data within the \texttt{ATAT} and \texttt{PhaseForge} frameworks, and subsequently optimized using \texttt{PyCalphad} and \texttt{ESPEI} to improve agreement with reported experimental observations.

%% file: 02_methods.tex
\section{Methods}
\subsection{Computation with PhaseForge}
To construct the CALPHAD model, we utilize our newly developed package, \texttt{PhaseForge}, which integrates the MLIPs into the ATAT framework. Special Quasirandom Structures (SQS)~\cite{zunger1990special} representing BCC and FCC phases, up to level~2, are adapted from the ATAT database. Structural relaxations and total-energy calculations of the SQS are performed using Grace~\cite{PhysRevX.14.021036} with the GRACE-2L-OMAT foundation model. Thermodynamic properties of unary end members, including energies and vibrational entropies, are obtained from the \texttt{PhaseForge} database, which is based on density functional theory (DFT) calculations conducted with VASP. For mechanically unstable phases---specifically FCC W and BCC Pt---inflection detection is applied, and the corresponding SGTE data are excluded from the CALPHAD assessment. Liquid-phase energies are calculated using a ternary-search method implemented in \texttt{PhaseForge}, with a temperature offset of \SI{50}{\kelvin}. In the CALPHAD-model fitting, binary interactions up to level~2 are considered, along with short-range order corrections for solid phases. The resulting TDB file, generated with \texttt{PhaseForge}, is subsequently imported into \texttt{PyCalphad} for Gibbs free energy calculations and phase diagram construction.

\subsection{Gradient-Based CALPHAD Model Parameter Optimization}

As detailed in~\cite{bocklund2019espei,kunselman2025analytical}, \texttt{ESPEI} employs a Maximum Likelihood Estimation (MLE) approach for CALPHAD model parameter optimization. In MLE, each data point used for the fitting of the model is treated as an observed value drawn from the distribution of a random variable associated with the conditions of the experiment or calculation that produced it. The likelihood function $L(\boldsymbol{\theta})$ is defined as the probability of drawing the entire dataset given the vector of model parameters $\boldsymbol{\theta}$. The goal is then to determine the set of parameters which maximizes $L(\boldsymbol{\theta})$, or equivalently, which makes the observed dataset most probable. To avoid floating point errors and facilitate easier differentiation, the log-likelihood $l(\boldsymbol{\theta})=\ln{[L(\boldsymbol{\theta})]}$ is the objective to be maximized. Under the assumptions that 1) all of the $n$ data points are independent and 2) the distributions of residuals for all data points can be modeled with a normal distribution with mean zero (because we want zero error to be most probable), the log-likelihood function takes the following form:

\begin{equation} \label{eq:log-likelihood}
l(\boldsymbol{\theta}) = \sum_{i=1}^n\left[\ln{\left(\frac{w_i}{\sigma_i\sqrt{2\pi}}\right)}  -\frac{1}{2}\left(\frac{w_iX_i}{\sigma_i}\right)^2\right],
\end{equation}

\noindent where $X_i$ is the residual of the data point $i$, $\sigma_i$ is a data-specific standard deviation that captures the uncertainty of $X_i$, and $w_i$ is a user-assigned weight factor that provides modelers with a knob to adjust the relative importance of each data point to the final solution. 

The gradient of \Cref{eq:log-likelihood} with respect to model parameters can thus be written as

\begin{equation}
    \frac{\partial l(\boldsymbol{\theta})}{\partial \boldsymbol{\theta}}=-\sum_{i=1}^n\left[\frac{w_i^2X_i}{\sigma_i^2}\frac{\partial X_i}{\partial \boldsymbol{\theta}}\right],
\end{equation}

\noindent where derivatives of the predicted portions of the model that require energy minimization of each $X_i$ are calculated using the Jansson derivative method in which the model parameters are treated as external conditions of the corresponding equilibrium calculations.

\begin{figure}[h!]
\centering
\includegraphics[width=\columnwidth]{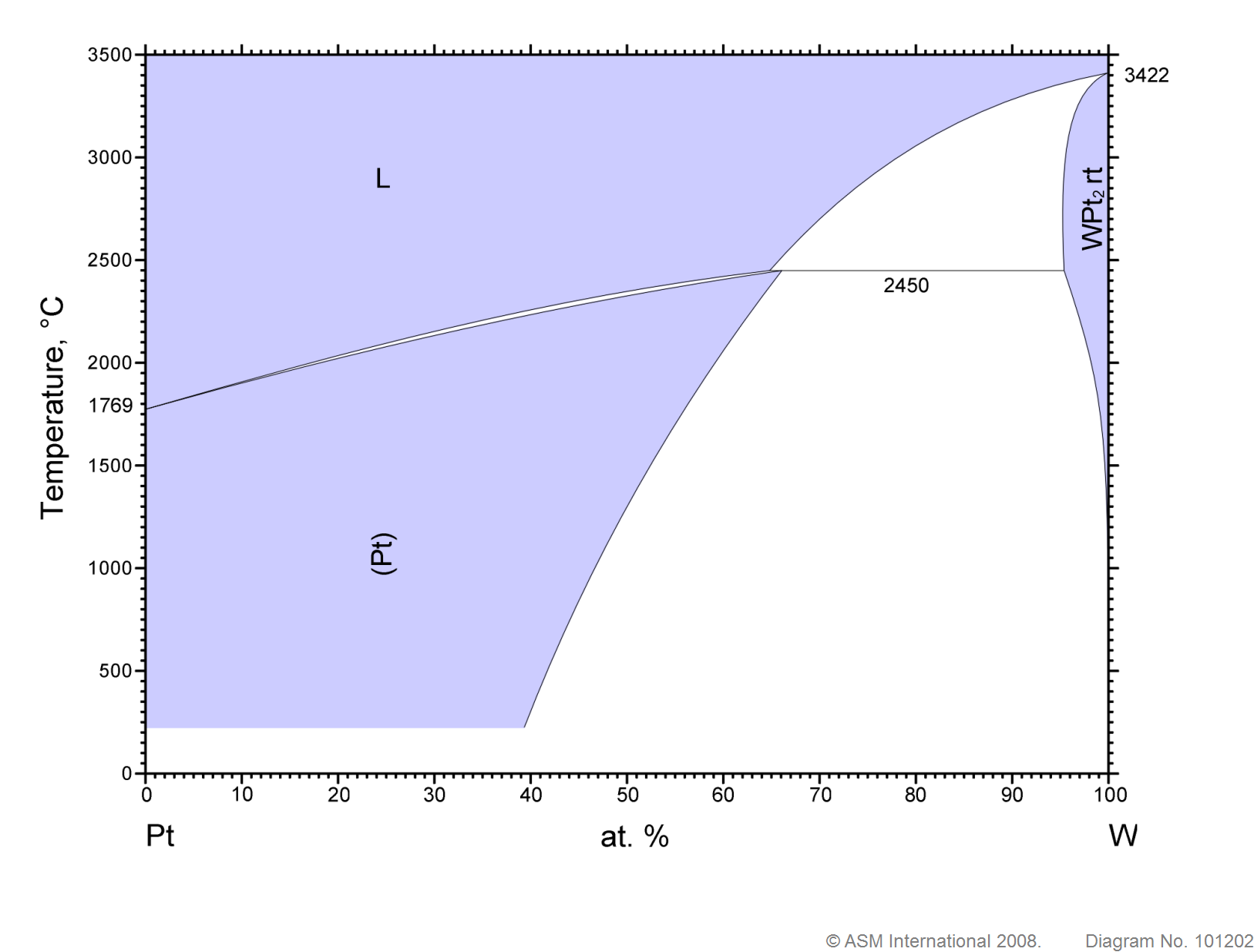}
\caption{Pt-W phase diagram reproduced with permission from JCPDS-International Centre for Diffraction Data \cite{asm}}
\label{fig:ASM_Pt-W}
\end{figure}

In this work, we use phase equilibria data extracted from the Pt-W phase diagram with number 101202 in the ASM Alloy Phase Diagram Database~\cite{asm} (see Fig. \ref{fig:ASM_Pt-W}) to build the log-likelihood function. The residuals for the phase equilibria data (referred to as zero phase fraction (ZPF) data in \texttt{ESPEI}) follow the residual driving force construction presented in~\cite{kunselman2025analytical,kunselman2024analytically,bocklund2021computational}:

\begin{equation} \label{eq:residual_driving_force}
    X^{zpf}_{i,\alpha} = \sum_A\bar{\mu}_Ax^\alpha_A-G^\alpha
\end{equation}

\noindent where $X^{zpf}_{i,\alpha}$ is the residual driving force of phase $\alpha$ for the data point $i$, $\bar{\mu}_A$ is the chemical potential of component $A$ corresponding to the target hyperplane, $x^\alpha_A$ is the reported vertex composition of component $A$ for phase $\alpha$, and $G^\alpha$ is the minimum Gibbs energy of phase $\alpha$ conditioned on the reported vertex composition. Unreported vertex compositions are estimated as described in~\cite{kunselman2025analytical}. For data points lying in or along the boundaries of single-phase regions with one specified vertex composition, the Case 2 target hyperplane construction described in~\cite{kunselman2024analytically} was used. For data points lying within two phase regions signifying the presence of two phases at a given temperature and overall composition but not specifying a tie line (i.e. both vertex compositions are unreported), the Case 1 hyperplane construction was employed. A visualization of the extracted phase equilibria data is provided in \Cref{fig:phase_diagram_tuned}.

To ensure that model parameters did not stray too far from the physically-informed values, mixing enthalpies of each phase were calculated from the \texttt{PhaseForge}-produced model across the composition space at \SI{298}{\kelvin} and added as non-equilibrium thermochemical data contributions to the log-likelihood function. In \texttt{ESPEI}, non-equilibrium thermochemical data simply denotes that internal phase degrees of freedom are fixed, and since all phases in this model have only one sublattice, providing the overall phase composition effectively specifies the internal configuration of each phase. Residuals for non-equilibrium data are simply the difference between the model-predicted and observed data values. Furthermore, since the phase configurations are fixed, no energy minimization calculations are necessary to calculate the model-predicted value. Accordingly, the analytic gradient functions of the residual can be easily obtained by partial differentiation of the target phase model with respect to the model parameters. In this way, additional equilibrium calculations are avoided, which helps to improve the computational efficiency of the workflow. All contributions to log-likelihood functions and their gradients are calculated using \texttt{ESPEI} pull request \# 268~\cite{espei_268}.

Similar to the approach in~\cite{kunselman2025analytical}, the gradient-based optimizer of choice in this work is \texttt{SciPy}'s conjugate-gradient implementation~\cite{2020SciPy-NMeth} with all hyperparameters set at default values. For preconditioning, we simply scale the large temperature-independent parameters by 1000 in order to put the magnitudes of their gradient components on par with those of the smaller temperature-dependent terms.

%% file: 03_results.tex
\section{Results and Discussions}
\begin{figure}[h!]
\centering
\includegraphics[width=\columnwidth]{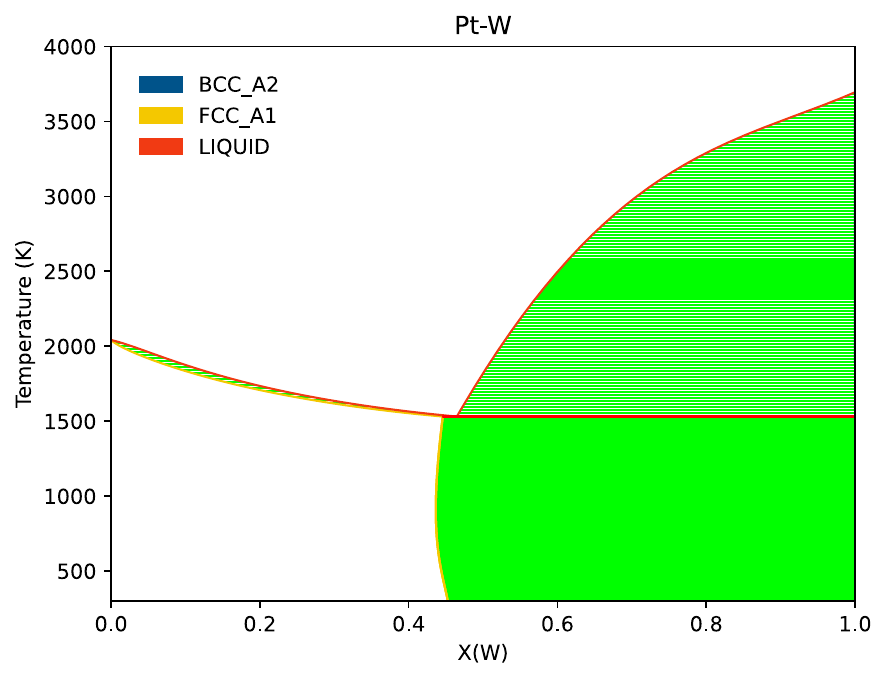}
\caption{Phase Diagram of the Pt-W system generated through ATAT and PhaseForge.}
\label{fig:phase_diagram_ATAT}
\end{figure}

In this work, we use the Pt–W binary system as an example to illustrate our workflow. First, we construct a CALPHAD model using MLIP-based thermodynamic data using \texttt{PhaseForge}, including BCC, FCC and liquid phases. The phase diagram is plotted using \texttt{PyCalphad}, as shown in~\Cref{fig:phase_diagram_ATAT}.

The calculated phase diagram correctly captures the stable FCC phase on the Pt-rich side and the BCC phase on the W-rich side. However, it significantly  underestimates the solubility of Pt in BCC-W, particularly at high temperatures, where experimental data indicate a non-negligible solubility over 5\%. Furthermore, the computed equilibrium temperature for BCC, FCC and liquid is over \SI{1000}{\kelvin} lower than experimental observations, which makes the eutectic reaction peritectic.

\begin{figure}[h!]
\centering
\includegraphics[width=\columnwidth]{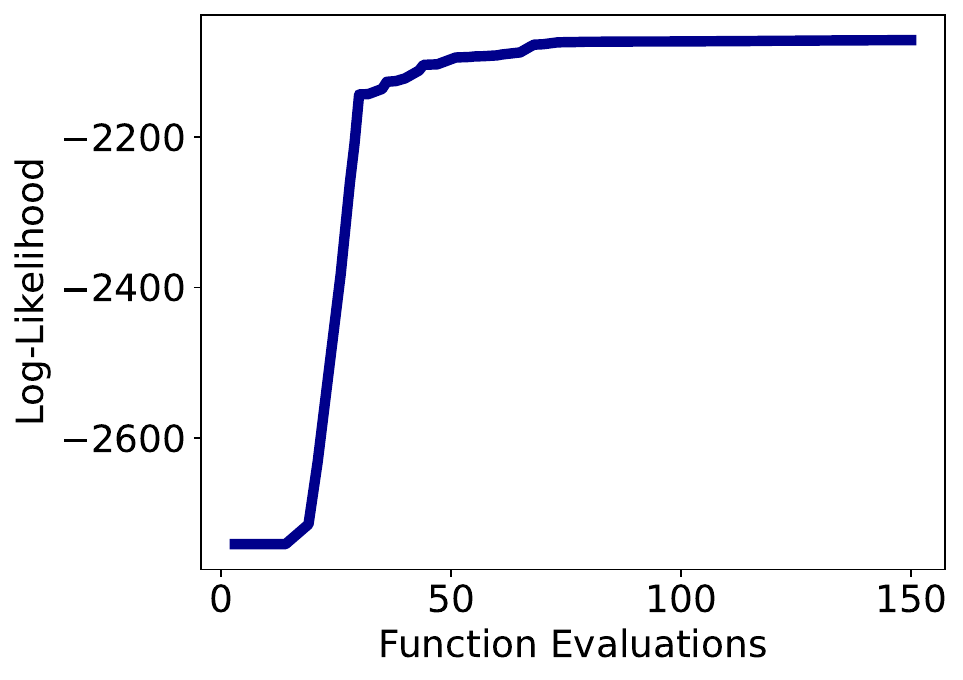}
\caption{Log-likelihood as a function of cumulative function evaluations.}
\label{fig:likelihood_convergence}
\end{figure}

\begin{figure}[h!]
\centering
\includegraphics[width=\columnwidth]{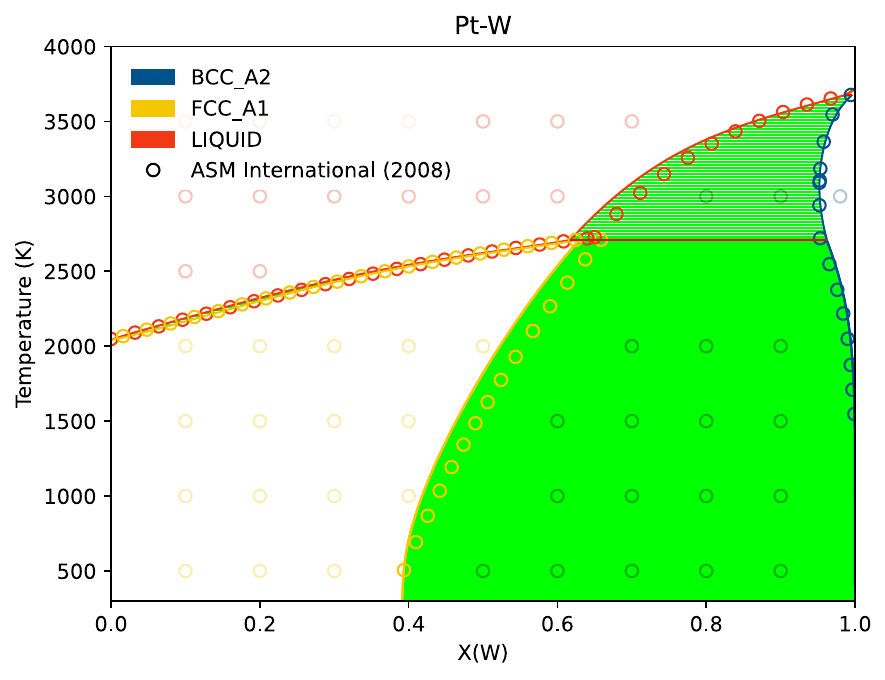}
\caption{Optimized phase diagram of the Pt-W system overlaid with phase equilibria data pulled from \cite{asm}. Gray circles represent the presence of two phases with no given vertex compositions.}
\label{fig:phase_diagram_tuned}
\end{figure}

Including non-SGTE unary end member parameters and excluding parameters associated with SGTE end member descriptions and short-range order contributions within interaction parameters, \texttt{PhaseForge} produced a model with sixteen tunable parameters. Before optimizing, an additional four parameters were added to provide linear temperature dependence for the zeroth- and first-order interaction parameters for the BCC and FCC phases. Fig. \ref{fig:likelihood_convergence} displays the progress of the optimization procedure with respect to cumulative log-likelihood function evaluations. The entire procedure required exactly \SI{150} function evaluations, but as shown in the plot, solutions with log-likelihoods within a tenth of a percent of the final solution were discovered after only \SI{80} function evaluations--orders of magnitude faster than the standard MCMC approach within ESPEI. The phase diagram of the final optimized model overlayed with the phase equilibria data is presented in Fig. \ref{fig:phase_diagram_tuned}, and we note excellent agreement between the data and the model predictions.

\begin{figure*}[h!]
\centering
\includegraphics[width=\linewidth]{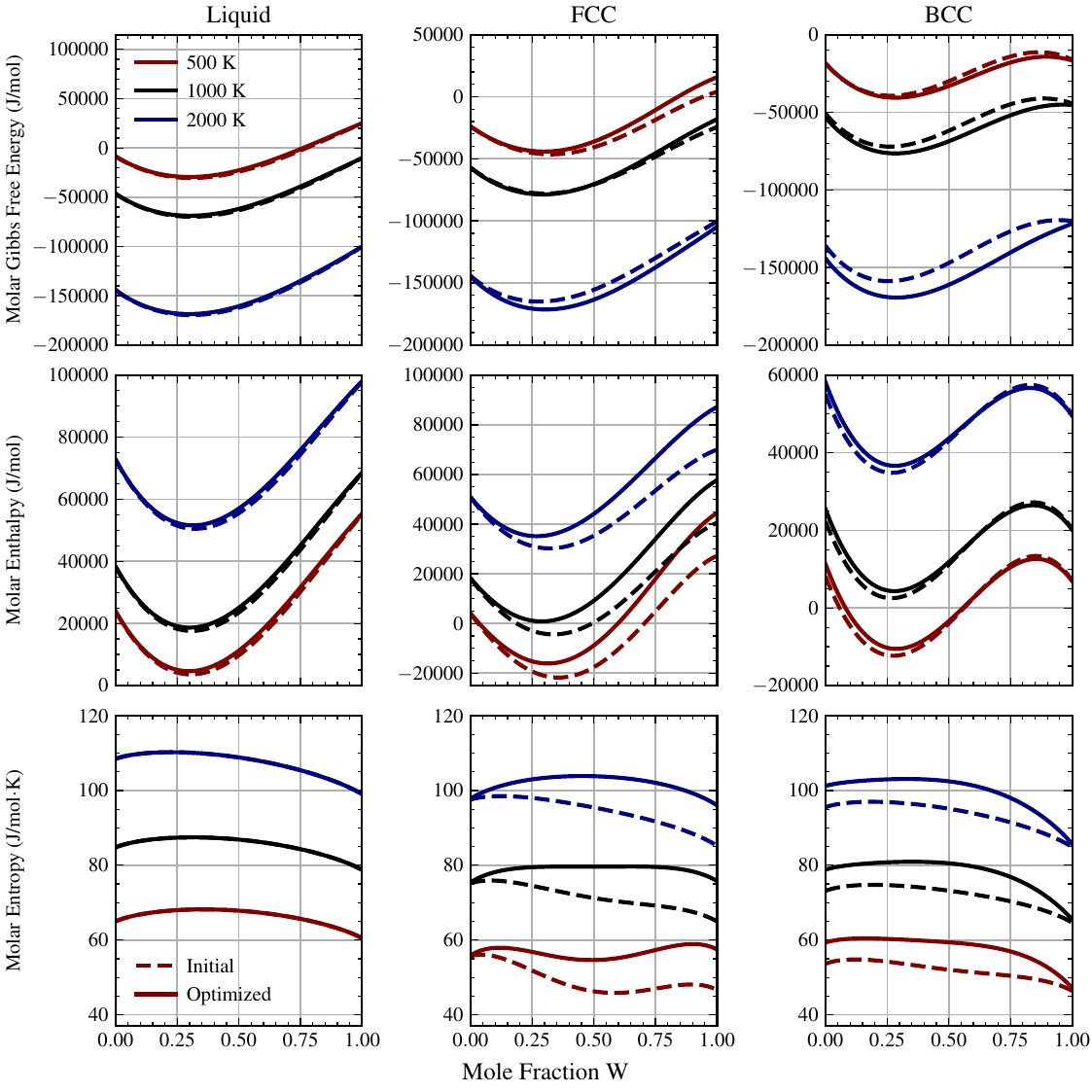}
\caption{Gibbs energy, enthalpy, and entropy per mole formula unit of each phase for the initial and optimized models at \SI{500}{\kelvin}, \SI{1000}{\kelvin} and \SI{2000}{\kelvin}.}
\label{fig:GM_HM_SM_plots}
\end{figure*}

In Figs. \ref{fig:GM_HM_SM_plots} we plot the Gibbs energy, enthalpy, and entropy per mole formula unit of each phase for the \texttt{PhaseForge}-generated and optimized models at temperatures of \SI{500}{\kelvin}, \SI{1000}{\kelvin}, and \SI{2000}{\kelvin}. From these plots and the corresponding phase diagrams, we see that relatively minor adjustments in the Gibbs energy descriptions through the optimization process resulted in massive improvements in phase equilibria predictions. The small changes in the Gibbs energy also give us confidence that the optimized model remains physically-informed. Further examination of Fig. \ref{fig:GM_HM_SM_plots} reveals that 1) the initial liquid phase description required very little fine-tuning and 2) small increases in enthalpy and entropy in the mechanically unstable FCC and BCC end member descriptions along with small modifications to interaction parameters was all that was necessary to produce an accurate phase diagram. For these mechanically unstable phases, their enthalpies and entropies are generally inaccessible from experiments directly, and extrapolations for the thermodynamic properties from different directions of stable phases might not match. We employed the inflection-detection method implemented in ATAT to obtain a quick estimation of these mechanically unstable phases. Although this method offers a practical solution for the unstable phases, the accuracy is limited and significantly affects the phase diagram predictions. Our subsequent parameter tuning successfully compensates for the inaccurate thermodynamic parameters of the mechanically unstable phases, showing that modest adjustments to their entropy and enthalpy can effectively bring the phase diagram back on the track. 

Lastly, we want to emphasize the computational efficiency of this workflow. Armed with only a previously-published phase diagram and local single-core computing resources, users can construct physically-informed, fine-tuned binary CALPHAD descriptions in less than one day. Thus, for users with access to super-computing resources, we foresee binary CALPHAD databases being built in a matter of days.

%% file: 04_conclusion.tex
\section{Conclusion}
In this brief case study, we demonstrated a computationally efficient, fully open-source workflow for constructing and optimizing physically-informed CALPHAD models. 

To further validate the thermodynamic parameters derived from this workflow, future work could explore the optimized descriptions of the mechanically unstable end members in systems such as Au-W or Cr-Pt. Or, the consistency of fitted binary interaction parameters could be examined in ternary systems such as Au-Pt-W. Looking further ahead, the computational efficiency afforded by gradient-informed optimizers could allow for the simultaneous fitting of unary unstable end member, binary interaction, and ternary interaction parameters in multi-component systems. Such an approach would allow experimental data from higher-order systems to influence the parameters of lower-order systems and build highly consistent hierarchical descriptions for systems with many components.

Furthermore, our work provides an efficient and systematic approach to parameter tuning in order to align the calculated phase diagram with experimental observations as closely as possible. In other words, it establishes a quantitative framework to evaluate CALPHAD modeling and the resulting phase diagrams based on current experimental knowledge. Traditionally, phase diagrams were often assessed only by visual comparison, since no rigorous quantitative criteria were available. For example, a diagram missing a stable phase or showing equilibrium temperatures deviating by hundreds of kelvin could be deemed inadequate, even if the calculated Gibbs free energy differed by only tens of joules per mole. By tuning CALPHAD parameters with reference to experimental data, we can quantitatively assess the discrepancy between calculated results and experimental truth, by the magnitude of parameter adjustments, or by the Gibbs free energies difference before and after tuning. Therefore, our framework would further enable a critical evaluation of all components used in any CALPHAD model construction process, including the ATAT/PhaseForge workflow, all parameters in the fitting process, DFT/MLIPs models used for calculations, the estimation of thermodynamic data of mechanically unstable phases, and even the quality of thermodynamic databases, from the perspective of calculated phase diagram, accurately, quantitatively, and efficiently.

%% file: 05_codeAvailability.tex
\section*{Code Availability}
The \texttt{PhaseForgePlus} code is publicly available at \url{https://github.com/dogusariturk/PhaseForgePlus}.

%% file: 06_acknowledgements.tex
\section*{Acknowledgements}

The authors acknowledge the support of the National Science Foundation, United States, through Grant No. 2119103 and OAC-1835690. We also acknowledge the support from the Army Research Office, United States, through Grant No. W911NF-22-2-0117. The authors also acknowledge the support from the U.S. Department of Energy (DOE) ARPA-E ULTIMATE Program through Project DE-AR0001427. Calculations were carried out at Texas A\&M High-Performance Research Computing (HPRC) Facility.